# Power Monitoring and Control for Large Scale projects: SKA, a case study

Domingos Barbosa*[a], João Paulo Barraca[a,b], Dalmiro Maia[c], Bruno Carvalho[d], Jorge Vieira[d], Paul Swart[e], Gerhard Le Roux[e], Swaminathan Natarajan[f], Arnold van Ardenne[g], Luis Seca[h]

[a]Instituto de Telecomunicações, Campus Universitário de Santiago, 3810-193 Aveiro, Portugal;
[b]Universidade de Aveiro, Campus Universitário de Santiago, 3810-193 Aveiro, Portugal;
[c]Faculdade de Ciências da Universidade do Porto, Rua do Campo Alegre, 4169-007 Porto, Portugal;
[d]Critical Software, Parque do Taveiro, Lote 49 3045-504, Coimbra, Portugal; [e]SKA Africa, 3rd Floor, The Park, Park Road, Pinelands, 7405, South Africa; [f]TCS Research, Tata Consultancy Services, IITM Research park, Chennai 600 113, India; ASTRON, Oude Hoogeveensedijk 4, 7991 PD Dwingeloo, The Netherlands; Center for Power and Energy Systems (CPES) INESC TEC - INESC Technology and Science Porto, Portugal

## ABSTRACT

Large sensor-based science infrastructures for radio astronomy like the SKA will be among the most intensive data-driven projects in the world, facing very high demanding computation, storage, management, and above all power demands. The geographically wide distribution of the SKA and and its associated processing requirements in the form of tailored High Performance Computing (HPC) facilities, require a Greener approach towards the Information and Communications Technologies (ICT) adopted for the data processing to enable operational compliance to potentially strict power budgets. Addressing the reduction of electricity costs, improve system power monitoring and the generation and management of electricity at system level is paramount to avoid future inefficiencies and higher costs and enable fulfillments of Key Science Cases. Here we outline major characteristics and innovation approaches to address power efficiency and long-term power sustainability for radio astronomy projects, focusing on Green ICT for science and Smart power monitoring and control.

## 1. INTRODUCTION

The Square Kilometre Array (SKA) is an international multipurpose next-generation radio interferometer, an Information and Communication Technology machine with thousands of antennas linked together to provide a collecting area of one square kilometer [1,5]. The SKA is the only global project in the European Strategy Forum of Research Infrastructures (ESFRI), with 10 Full members (Australia, Canada, China, Germany, Italy, New Zealand, South Africa, Sweden, The Netherlands and the United Kingdom) and an Associated member (India). It further involves more than 67 organizations in 20 countries, and counts with world-leading Information, Computing and Telecommunication (ICT) industrial partners. The SKA will be built by Phases in the Southern Hemisphere, first with Phase 1 in South Africa and Australia (SKA1, starting in 2018), spreading in Phase 2 (SKA2, starting in 2022) to SKA African Partners - Botswana, Ghana, Kenya, Zambia, Madagascar, Mauritius, Mozambique, Namibia - and Australia/New Zealand. SKA will consist of a central core of ~200 km diameter, with 3 spiral arms of cables connecting nodes of antennas spreading over sparse territories in several countries up to 3000km distances. Since the SKA will continuously scan the sky, it will present a strong need for quality of service of its IT infrastructure to achieve high operational availability. SKA presents the opportunity for a combination of low power computing, efficient data storage, local data services, inclusion of newer Smart Grid power management, and inclusion of local energy sources, including potential Renewable Energies.

*dbarbosa@av.it.pt; phone +351 234 377 900; fax +351 234 700 901

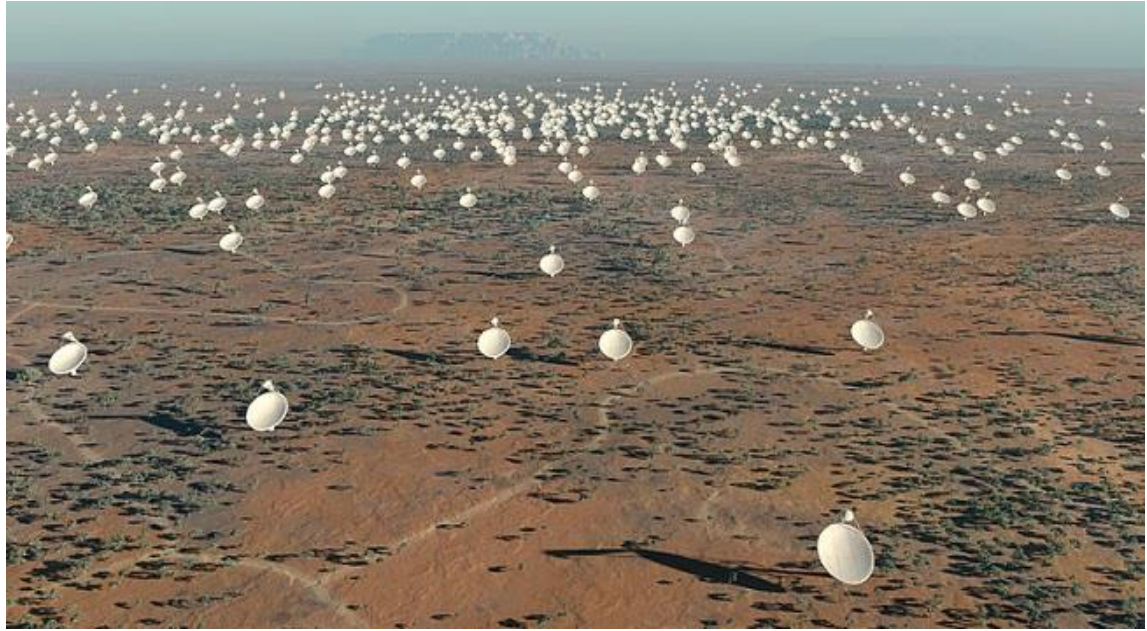

Figure 1 - An artist vision of the SKA Core site, with some of the projected 3000 15-meter parabolic dishes. From http://www.skatelescope.org.

The SKA is broken down in to various elements that will form the final SKA Observatory managed by an international consortium comprising several world leading experts in their respective fields. We should note that SKA will be in fact constituted by two Telescopes, SKA1-Low (Australia) and SKA1-MID (South Africa), each one providing their own Elements subsystems. Table 1 provides the characteristics of each SKA Phase 1 telescope (hence SKA1). A distinct Element product tree will be designed for each Telescope during the Pre-Construction Phase. These products will be based on a common architecture and design as far as possible, except for the Dish products, to be deployed in South Africa and the Low Frequency Arrays products to be deployed in Australia.

|  | **SKA1_LOW (Australia)** | **SKA1_MID (South Africa)** |
|---|---|---|
| **Sensors type** | 130 000 dipoles | 197 Dishes (including 64 MeerKAT) |
| **Frequency range** | 50-350 MHz | 0.45-15 GHz |
| **Collecting Area** | 0.4 Km$^2$ | 32 000m$^2$ |
| **Max baseline** | 65 Km (between stations) | 150 Km |
| **Raw Data Output** | 0.49 Zettabyte/year | 122 Exabyte/year |
| **Science Archive** | 128 Petabyte/year | 1.1 Exabyte/year |

Table 1 - SKA Phase1 Telescopes Broad Characteristics

Like any major large-scale astronomy projects, installed usually in remote locations, the associated (power hungry) data processing location is conditioned by the experiment, and not by the computational facilities, resulting in far from optimal efficiency, higher capital expenditure (CAPEX) and higher operational expenditure (OPEX). Addressing both the reduction of electricity costs and the generation and management of electricity at system level is paramount to avoid future inefficiencies and higher costs. Remote locations also imply development of customized supply grid that may be built in phases, preceding the deployment of the experiments. This means usually power caps may be imposed with consequences on Key Science prioritization. Phasing of projects also alleviates concerns with Infrastructure and power budgets: it is easier to aggregate sensors (antennas) and upgrade processing facilities once Infrastructure including power is expanded to cope with the planning of Science Operations. However, fulfillment of certain Science Cases including observations of Transient or other Virtual Observatory (VO) triggered observations may produce sudden peak power loads.

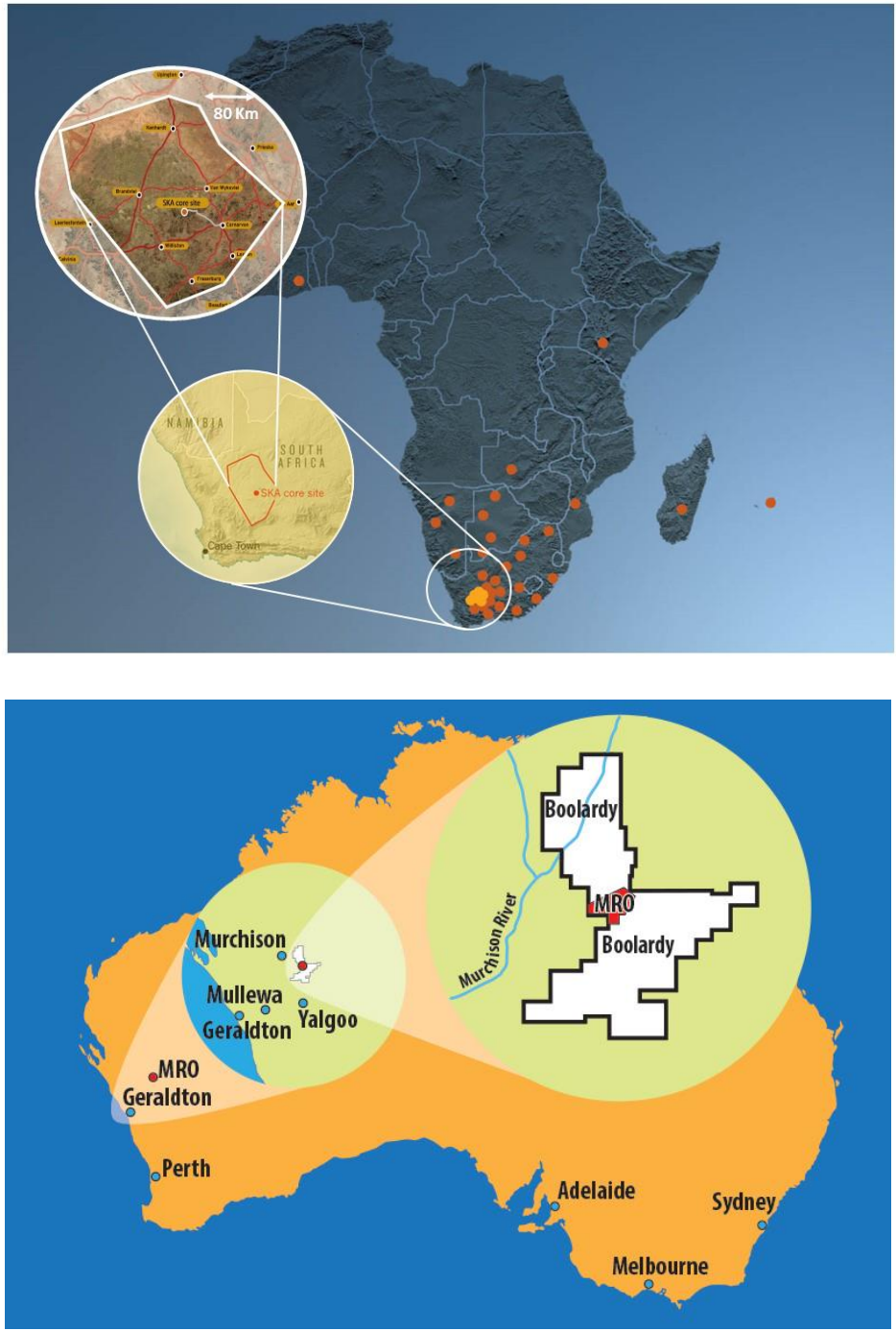

Figure 2 -Top: the SKA1 Radio Quiet Zone, in the Karoo, South Africa. Also shown are the planned SKA2 locations in the SKA Africa Partner Countries. Bottom: The SKA1 Radio Quiet Zone of the Murchinson Shire of Western Australia (@ SKA Australia).

As an example, these unexpected, yet extremely relevant astronomical events will not only require precise power metering, but also the capacity to manage overall system dispatch, considering technical and economic constraints. So it becomes clear that this would imply that to deal with these operational modes, it is necessary to forecast load and generation but also to remotely control some of the telescope components to not compromise the capacity to register these phenomena. Hence, to be able to support an adequate operation for this system, we should plan in detail the dimension of the local renewable based generation, the storage needs and at the same time some flexibility from demand, by curtailing unnecessary services/appliances whenever possible. The local electrical network should also have the capacity to adapt to the different operation scenarios, namely by reconfiguration actions and also by multi-temporal management of distributed resources.

## 2. POWERING LARGE SCALE PROJECTS

Energy Sustainability of large-scale scientific infrastructures and the related control of OPEX means inclusion of the impact of power provision costs and associated their carbon footprint into the respective development path and lifetimes [2]. Additionally, the ESFRI has indicated that it is paramount that a multitude of test facilities and Research Infrastructures should lead the world in the efficient use of energy, promote new renewable forms of energy, and develop low carbon emission technologies, to be adopted as part of a future Strategic Energy Technology Plan [2]. Radio astronomy projects will be among the most data-intense and power hungry projects. Recent experiences with SKA precursors and pathfinders like ASKAP, MeerKAT and LOFAR, MWA reveal that an important part of the life cycle cost of these large-scale radio astronomy projects will be power consumption [6],[7]. As an example, a 30-meter radio telescope requires approximately during operation1GWh for a typical 6h VLBI observation experiment, enough to power a small village, while new infrastructures based on Aperture Arrays, promising huge sky survey speeds, may require even more, based on estimated digital processing needs [7]. Among the large scale well-known science infrastructures facilities, SKA will set the highest constrains on power consumption and availability, surpassing current figure considerably as can be perceived by Table 2. This is due to geographical spread of the SKA sensor network, ancillary facilities deployed in remote locations in the Karoo savanna in South Africa and the Western Australia desert

| SKA Phase 1&2 | South Africa | Australia |
|---|---|---|
| Sparse Arrays | | 3.36 MW |
| Mid Dishes | 2.5MW | |
| Survey Dishes | | 1.2MW |
| On-site Computing | 4.7MW | 1.32MW |
| Totals/site | 5.7MW | 4.8MW |
| SKA Phase2 incl. Dense Arrays | >40MW (SKA Phase 2 configuration not known yet) | |
| Off-site Computing | ~20-40MW (SKA Phase 2 configuration not known yet) | |

Table 2 - SKA Energy Budgets

Addressing both the reduction of electricity costs and the generation and management of electricity is paramount to avoid future inefficiencies and higher costs. For instance, the Atacama Large Millimeter Array (ALMA) interferometer and the Very Large Telescope (VLT) in the Chilean Andes are powered from diesel generators, leading the European Southern Observatory (ESO) to consider potential inclusion of local greener energy sources to its Very Large Telescope (VLT) facilities in Paranal [8]. For VLT, electrical power is produced in off-grid mode using a combination of efficient multi-fuel turbine generators (2.6MWe at the site) that can use fuel sources with lower carbon footprint like natural gas, or Liquefied Petroleum Gas (LPG), combined with diesel generators connected to a 10kV power grid. However, fossil based fuels experience strong market fluctuations, with the overall long term trend showing a steep price increase. In a recent study, it was pinpointed the impacts of the increasing costs in electricity provision in Chile : between 2003 and 2010 the price rose by 7% per year according to statistics from the Organization for Economic Cooperation and Development (OECD) [6, 8]. Therefore, the ALMA permanent power system plant, capable of providing up to 7MW peak in "island - mode" is already prepared to connect to a renewable power plant, and the European Extremely Large Telescope (E-ELT) might include options for local renewable provision when market options in Chile make these technologies economically accessible [8]. Hence, from the pure power provision point of view and associated control of operational costs, fossil fuel price fluctuations and longer term availability and associated price rises represent a challenge in terms of planning a suitable energy mix supply, in particular for remotely located infrastructures. Overall, the main characteristic concerning the SKA power system can summarized as:

- Many Antennas nodes are far away from civilization centers and power grid in climates with high thermal amplitudes.

- Exquisite control of Radio Frequency Interference and EMI from Power systems is needed, since RFI would impair the radio telescope sensitivity.

- Different Power requirements over large distances.

- Continuous operation (meaning 24/7 availability) for sky surveying points out that some storage capabilities may be required, and power supply for night operations must be carefully considered.

- Power balance and control: control of large power peaks, for operation, cooling, computing and telescope management and monitoring, as a mean to maximize the integration of renewables based energy sources.

- Scalability: the power infrastructure should scale from SKA Phase 1 to the later, more extended, and more power demanding Phase2 (see Table **2**).

## 3. THE IMPACT OF GREENER ICT

The biggest computing challenge within radio astronomy lies within the architecture of the correlator of big synthesis radio telescopes and the second tier processing and storage infrastructures. The correlator processes the data streams arising from the large number of antenna elements of say, with N>1000 antennas. The optimum architecture is planned to minimize power consumption as much as possible by following several approaches: minimizing I/O (storage media, and network interconnects) and memory operations, implying preference for a matrix structure over a pipeline structure and avoiding the use of memory banks. For instance, the ALMA correlator selected for its core design the StratixII 90nm technology based on considerations on power dissipation and logic resources while much lower power technologies are available now. The SKA, under the Central Signal Processor Element Consortium, is currently developing design concepts in a power efficient way for design for N>2000 and over 1 GHz frequency bandwidth, based on Application-specific integrated circuits (ASICs) fabricated in a ~20nm CMOS process, still better than 20nm for FPGAS with low power considerations. Also, the integrated circuit (IC) design that performs digital cross-correlations for arbitrarily many antennas in a power-efficient way using intrinsically low-power architecture in which the movement of data between devices is minimized. Excluding antenna data pre-processing, the SKA correlator is estimated to consume less than 100 kW [3].

Hence it is expected a great advance on the low-power processing capabilities of SKA correlators. After data is integrated by the correlator and further processed to create calibrated data, it must be stored in a permanent media, such as the case of massive Storage Area Networks (SANs), relying in rotational technologies such as hard disks.

| Telescope | Status | Technology | Design Freeze Year | N of elements | Power efficiency (pJoules) |
|---|---|---|---|---|---|
| VLA | Obsolete | ASIC | 1975 | 27 | 171k |
| JVLA | Existing | ASIC 130 nm | 2005 | 32 | 4270 |
| ALMA | Existing | ASIC 250 nm | 2002 | 64 | 992 |
| LEDA | Existing | GPU 28 nm | 2011 | 256 | 977 |
| CHIME | Existing | GPU 28 nm | 2013 | 128 | 769 |
| SKA1-Low | Proposed | FPGA 16 nm | 2017 | 512 | 74 |
| SKA1-Low | | ASIC 32 nm | 2015 | 512 | 4.8 |
| SKA1-MID | Proposed | FPGA 16 nm | 2017 | 197 | 103 |
| SKA1-MID | | ASIC 32nm | 2015 | 197 | 12.7 |
| SKA2 | Planned | TBD | 2021 | >2000 | TBD |

Table 3 - Power efficiency of Radiotelescope Correlators. From D'Addario and Wang (2016)

ALMA can output several TeraBytes of data per project that must be stored, and the future SKA infrastructure is expected to produce closer to an Exabyte/day of raw information, prior to further processing and data reduction (see Table 1). All these data must be made available in large facilities for further reduction by researchers (eg, using CASA [13] or other parallelized data handling software pipeline requiring most probably a high degree of automatisms). Due to the amount of information, and the costs of transmitting data through long distance optical links, it is vital the use of computation facilities located in close proximity to the source of information, but also close to researchers, in order to reduce latency and cost of the post-analysis process. Hence at SKA, most of the compute power will be located in Central Processing Facilities, properly shielded, in the vicinity of the Radio Quiet Zones in the Karoo and Western Australia.

The typical approach is to create computational behemoths capable of handling the entire operation of the instruments, storage, and frequently further processing of the data produced. However, lessons learned from similar large infrastructures show firsts years' operations to have frequent interruptions caused by detection of erroneous or unexpected behavior, requiring further tuning, or even due to integration of components arising from phasing of project deployment. Even after entering into its normal operational status, instruments are, among other factors, affected by maintenance downtime, and also by weather conditions limiting observations. As an example, according to the ALMA cycle 0 report, over the course of 9 months (total of ~6500 hours), the instrument was allocated for 2724 hours of observation time, and this resulted in 38% (1034 hours) of successful observation [9]. This results in a considerable efficiency loss, considering all the processing infrastructure that must be available, independent of the observation status. Although we believe the initial processing must be done close to the location of the sensors, data processing should be shared or co-located as much as possible to other already existing infrastructures, exploiting time multiplexing as a way of increasing power efficiency. Moreover, further offline reduction methods can be improved as they currently typically use dedicated hardware and facilities, which are only used after a successful observation is obtained, further increasing the OPEX and the carbon footprint of science.

From this, it is clear SKA requires signal and data processing capacities exceeding current state-of-the art technologies. The DOME project [18,19] is a 5-year collaboration between ASTRON, South Africa and IBM, aimed at developing emerging technologies for large-scale and efficient (green) exascale computing, data transport, storage, and streaming processing. Based on experience gained with a retrospective analysis of LOFAR, the DOME team analyzed the compute and power requirements of the telescope concepts for the first phase of the SKA [12]. These initial estimates indicate that the power requirements are challenging (up to order ten peta operations per second (OPS) in the station processing and correlation), but especially the post correlation processing (order 100 peta OPS to exa OPS) is dominating the power consumption [12, 18]. The study also poses mitigation strategies, such as developing more efficient algorithms, fine-tuning the calibration and imaging processing parameters, and phased-implementation of novel accelerator technologies.

From the perspective of Green Computation, there are several aspects that have been tackled in order to increase the efficiency and decrease operational costs (OPEX) of current infrastructures such as: location, infrastructure reuse, equipment selection (servers, racks, networking), and cooling parameters. Recently, the reuse of devices reaching their end-of-life has also been addressed as a way to reduce the ecological footprint of a given system. Ultimately, if considering the operational stage of a datacenter, the most common metrics for evaluating the efficiency of a computational infrastructure are FLOPS per Watt (F/W) and Power User Efficiency (PUE), where PUE= (Total Facility Energy/ Information Equipment Energy). An ideal PUE value would be 1, whereas state-of-art is already PUE $\leq$ 1.2 for some greener large datacenters. Most of these metrics can also be applied to large Science Infrastructures. In addition, some tasks may be off-loaded to public clouds having lower PUE values.

Location is a major aspect driving the development of a large computational cloud facility. Ideally, a data center should be placed next to a power source so that the price is minimum, and losses in the power grid are minimized (est. 17% is lost in the power grid [25]). For projects with grid supply difficulties and with ecological aspects, as it is common to observe, presence of water dams, wind turbines and solar panels may be considered, provided strict compliance the Radio Quiet Zone maximum allowed interference requirements. Moreover, if possible, to decrease OPEX there must be interconnectivity to the global Internet through multiple providers although in remote locations like the Karoo and Western Australia, this job may be provided only by the local National Research and Education Networks (NRENs) like the TENET/SANReN (South Africa) and AARNet (Australia). Climate and geography also play an important role with great impact in temperature control, and overall security of the infrastructure, although at the desertic SKA locations, free cooling can certainly be a problem due to the absence of water planes (rivers and oceans) or wind.

However, for most large science facilities location is conditioned by the experiment, and not by the computational facilities, which results in far from optimal efficiency, higher capital expenditure (CAPEX) and higher OPEX. As an example, ALMA, with its correlator located in the middle of the Atacama Desert, at an altitude of 5km, far from power sources, and with a thin atmosphere, presents serious engineering challenges even for keeping basic operation, and just without addressing efficiency concerns. Infrastructure reuse is another important aspect that is always considered, and at multiple levels.

In the area of computing and Internet service provisioning, it is possible to increase the usage rate of computational resources (servers), by exploring virtualization and service-oriented technologies, mostly due to the intermittent resource consumption pattern shown by almost any application or service. By combining multiple, unrelated services in the same hardware resources, processing cycles can be multiplexed, ensuring that overcapacity is reduced to a minimum. Using this technique, servers are optimized and redesigned to become highly power efficient. As a practical example, some commercial cloud providers exploit these properties by providing spot pricing for their resources, according to the laws of demand and supply. In this aspect, Cloud Computing technologies have emerged as a promising Green ICT solution, which can be exploited by Big Data Centers and Science Organizations [6]-[8], thus addressing also the management and power concerns of large scale science infrastructures. Hence, the concepts of Infrastructure as a Service (IaaS), Platform as a Service (PaaS) or even the lately developed Software as a Service (SaaS), can provide abstraction from physical compute infrastructure and potentiate data center operators to trim energy costs and reduce carbon emission. Furthermore, software development sourcing on the emergent DevOps ideas from the telco/IT sectors promote agile resource management, automating the process of software delivery and infrastructure through microservice architectural delivery with high modularity. DevOps practice ensures a set of Architecturally Significant Requirements (ASRs) such as deployability, modifiability, testability, and monitorability. These ASRs require a high priority, allowing the architecture of an individual service to emerge through continuous refactoring, hence reducing the need for a big upfront design, reconfiguration of physical infrastructure underneath and reducing the time to market introduction of well-developed software services via frequent software releases early and continuously.

## 4. THE MONITORING TECHNOLOGIES: SMART GRIDS

The specificity of demand, namely by a significant amount of scenarios with high levels of power requirements, support the implementation of the smart grid paradigm for the local distribution network. A truly smart grid will rely on adequate monitoring, communication and control over existing network, including flexibility coming form generation, storage and demand. This flexibility bears in mind that load and generation forecast are also included, being each of the SKA sites run by a Supervisory Control and Data Acquisition (SCADA) Distribution Management System (SCADA-DMS) that will support a more efficient, reliable and sustainable operation of each of the sites. Power monitoring of antennas and ancillary systems, Correlators, HPC facilities or related data center tiered systems must include advanced remote metering technologies, efficient distribution automation and power Network Operation Centers (NOC).

SCADA protocols are designed to be very compact, and we do expect SCADA-DMS system filtered information to be provided by INFRA element to the Telescope Manager (The Operational, Monitoring and Control Element of the SKA). SCADA also improve reliability, increase resource utilization and contribute to OPEX reduction.

In a typical configuration, power substations are controlled and monitored in real time by a Programmable Logic Controller (PLC) and by power-specialized devices like circuit breakers and power monitors. PLCs and the associated devices communicate data to SCADA node located at the substation. The links between the substation PCs and the central station PCs are generally Ethernet-based and may be implemented via the SKA Non-Science Data Network intranet. In some cases, the information could even use private versions of cloud computing. SCADA systems feature built-in redundancy and backup systems to provide adequate reliability, and can be deliver much faster-acting automated control that can greatly benefits utilities and consumers, in this case can benefit Large Scale Infrastructures like SKA. Capabilities include systemic problem detection with alarm handling and trigger adjustments and corrections, often preventing an outage when more serious problems may arise. These SCADA-DMS capabilities largely benefit extended sensor networks since they enable maintenance teams to identify the exact location of outage or any other major critical problems that may affect the Telescopes performance, and thus significantly increase the power stability and the speed of power restoration in the case of an outage via fast rerouting o power for unaffected regions without the need for maintenance visual inspections.

Besides allowing system operators to use powerful trending capabilities to forecast future problems, SCADA system allow storage of data for profiling the quality of power supply properties (voltage levels, power factors, other system parameters) across grid and hence across any SKA component subsystem. If economically viable, inclusion of any local power source generation (like renewable sources) can make power quality more difficult to achieve, thus requiring more automated responses since power supplied to the distribution system would come from multiple sources in addition to the large base-load power stations.

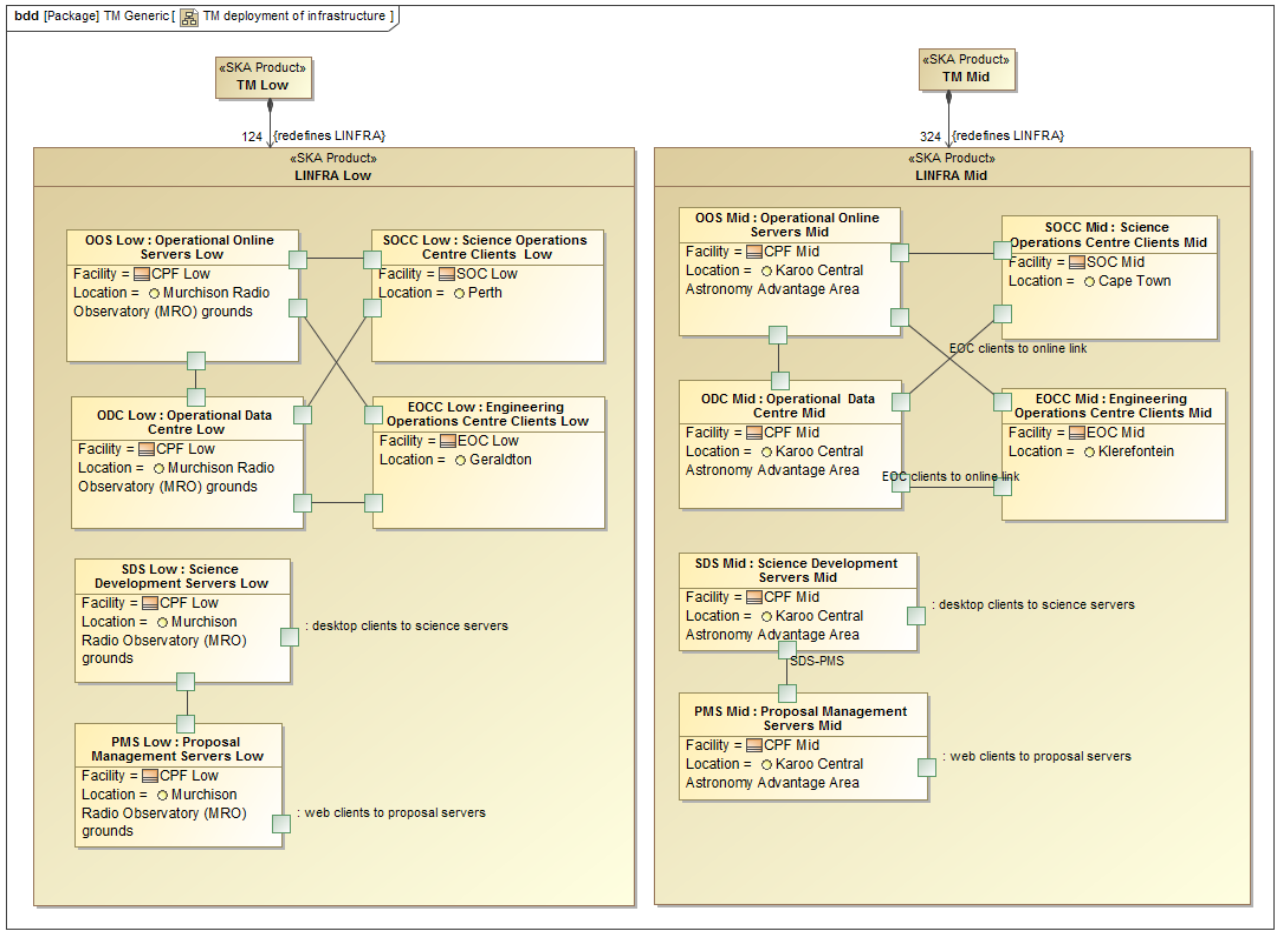

Figure 3 And example of infrastructure co-location: the Telecope Manager Infrastructure Deployment, itself responsible for the monitoring and operations of the Telescopes, to be collocated with the Central processing facility in the vicinity of the core-sites. Although most of the LINFRA will be co-located in the Central Processing Facility (CPF) of each telescope, several instances will also be deployed in the Science Operations Centres (SOC) and the Engineering Operation Centers (EOC).

## 5. CONCLUSIONS

Here we outline the major characteristics and innovation approaches to address power efficiency, life cycle impact, and long-term power sustainability for radio astronomy projects like the SKA. The current trends in the area of Green ICT, embodied in the Cloud Computing technologies and DevOps software ideas are influencing the design of compute infrastructures like Data centers and will be key in improving the power operational cost of SKA. The inclusion of Smart Grid technologies will raise greater efficiency and will provide capabilities including detailed power forecasts, improved service reliability, more efficient power asset management, and better operational planning. Power monitoring of antennas and ancillary systems, Correlators, HPC facilities or related data centre tiered systems must include advanced metering technologies, efficient distribution automation and Network Operation Centres (NOC).

# 6. ACKNOWLEDGMENTS

DB and JPB acknowledge support from FCT through national funds and when applicable co-funded by FEDER – PT2020 partnership agreement under the project UID/EEA/50008/2013. DM acknowledge support from FCUP. The Portuguese team contributed through the ENGAGE SKA Research Infrastructure.